\documentclass[12pt]{article}

\usepackage[margin=1in,nohead,a4paper,dvips]{geometry}

\usepackage{amsfonts,amsmath}
\allowdisplaybreaks

\usepackage[noBBpl,slantedGreek]{mathpazo}

\def\bbC{\mathbb C}

\def\HXXZ{H_{\mathrm{XXZ}}}
\def\rmd{\mathrm d}
\def\rmi{\mathrm i}

\begin{document}

\title{Bethe roots and refined enumeration of alternating-sign matrices}
\author{A.~V.~Razumov, Yu.~G.~Stroganov\\
\small \it Institute for High Energy Physics\\[-.5em]
\small \it 142281 Protvino, Moscow region, Russia}
\date{}

\maketitle

\begin{abstract}
The properties of the most probable ground state candidate for the XXZ spin chain with the anisotropy parameter equal to $-1/2$ and an odd number of sites is considered. Some linear combinations of the components of the considered state, divided by the maximal component, coincide with the elementary symmetric polynomials in the corresponding Bethe roots. It is proved that those polynomials are equal to the numbers providing the refined enumeration of the alternating-sign matrices of order $M + 1$ divided by the total number of the alternating-sign matrices of order $M$, for the chain of length $2 M + 1$.
\end{abstract}

\section{Introduction}

The XXZ spin chain is a system consisting of $N$ spin one-half particles arranged in a linear structure. Here each particle interacts with its nearest neighbors only. The state space $V$ of the XXZ model is the tensor product of $N$ copies of the state space of a single spin one-half particle, which is isomorphic to the complex two dimensional coordinate space $\bbC^2$. Denote the vectors of the standard basis of the vector space $\bbC^2$ as $| + \rangle$ and $| - \rangle$, so that
\[
| + \rangle = \begin{pmatrix} 1 \\ 0 \end{pmatrix}, \qquad | - \rangle = \begin{pmatrix} 0 \\ 1 \end{pmatrix}.
\]
The standard basis of the state space of the XXZ model is formed by the vectors
\[
| \mu_1, \mu_2, \ldots, \mu_N \rangle = {| \mu_1 \rangle} \otimes {| \mu_2 \rangle} \otimes \ldots \otimes | \mu_N \rangle,
\]
where the variables $\mu_1, \mu_2, \ldots, \mu_N$ take the values $+$ and $-$. Sometimes it is convenient to assume that such variables take the values $+1$ and $-1$. The used interpretation is always clear from the context.

Denote by $\sigma^x$, $\sigma^y$ and $\sigma^z$ the usual Pauli matrices and the corresponding operators in $\bbC^2$. For any $n = 1, 2, \ldots, N$ define
\[
\sigma_n^i = {\underbrace{I \otimes \cdots \otimes I}_{n-1}} \otimes \sigma^i \otimes {\underbrace{I \otimes \ldots \otimes I}_{N-n}}, \qquad i = x, y, z.
\]
Here we denote by $I$ the identity operator in $\bbC^2$. Now we are able to write the Hamiltonian of the XXZ spin chain. It has the form
\begin{equation}
\HXXZ(\Delta) = - \frac{1}{2} \sum_{j=1}^{N} \left[ \sigma_n^{x} \sigma_{n+1}^{x}   + \sigma_n^{y} \sigma_{n+1}^{y} + \Delta \, \sigma_n^z \sigma_{n+1}^z \right],
\label{H1}
\end{equation}
where $\Delta$ is the anisotropy parameter. In the present paper we use the periodic boundary conditions.

As for any quantum mechanical system, the main task related to XXZ spin chain is to find the eigenvalues and eigenvectors of the Hamiltonian. Before discussing the methods used to find them, let us make two general remarks. Note that the operator
\[
\Sigma = \sum_{n=1}^N \sigma_n^z
\]
commutes with $\HXXZ(\Delta)$. Therefore, one can look for the eigenvectors of the Hamiltonian $\HXXZ(\Delta)$, restricting oneself to a subspace spanned by the eigenvectors of $\Sigma$ with a fixed eigenvalue. It is clear that
\[
\Sigma \, | \mu_1, \ldots, \mu_N \rangle = \left( \sum_{n=1}^N \mu_n \right) | \mu_1, \ldots, \mu_N \rangle = (N_+ - N_-) | \mu_1, \ldots, \mu_N \rangle,
\]
where $N_+$ and $N_-$ are the numbers of pluses and minuses,respectively, in the set formed by $\mu_1, \ldots, \mu_N$. Hence we can look for the eigenvectors of the Hamiltonian $\HXXZ(\Delta)$ in a subspace spanned by the basis vectors with a fixed number of pluses, or minuses. Considering such subspaces, we will denote by $K$ the number of minuses. 

Note here that the minimal eigenvalue in a subspace with a fixed $K$ is non-degenerate \cite[Theorem 4]{YanYan66}.

It is also not difficult to get convinced that the Hamiltonian $\HXXZ(\Delta)$ commutes with the operator
\[
P = \prod_{n=1}^N \sigma_n^x.
\]
One easily obtains that
\[
P \, | \mu_1, \ldots, \mu_N \rangle = | {- \mu_1}, \ldots, - \mu_N \rangle.
\]
Therefore, each eigenvector $| \Psi \rangle$ in the subspace with $K$ minuses has the companion $P \, | \Psi \rangle$ in the subspace spanned by the basis vectors with $N - K$ minuses. The both eigenvectors certainly correspond to the same eigenvalue. In the case of an even $N = 2M$ and $K = M$ two eigenvectors connected by the operator $P$ belong to the same subspace and can coincide.

It is clear that the subspaces with $K = 0$ and $K = N$ are one dimensional. Therefore, the corresponding basis vectors must be eigenvectors of $\HXXZ(\Delta)$. Thus, we know at least two eigenvectors of $\HXXZ(\Delta)$. These eigenvectors are ground states for the XXZ spin chain only for the case $\Delta \ge 1$. They do not depend on $\Delta$, and the corresponding correlation functions are not quite interesting from the physical point of view. 

For $\Delta < 1$ and an even $N = 2M$ the ground state of the XXZ spin chain is unique and belongs to the subspace with $K = M$, see, for example, the monographs \cite{Bax82,IzeKorBog93}. For $\Delta < 1$ and an odd $N = 2M + 1$ we have two ground states connected by the operator $P$, one in the sector with $K = M$, and the other one in the subspace with $K = M + 1$.

In general the eigenvectors and eigenvalues of the Hamiltonian $\HXXZ(\Delta)$ can be found with the help of the famous Bethe ansatz \cite{Bet31}. Unfortunately, this method provides us with the expressions of the eigenvectors and eigenvalues via the roots of the Bethe equations, which are not known usually. 

Another method, partially equivalent to the Bethe ansatz, is based on the relation of the Hamiltonian $\HXXZ(\Delta)$ and the transfer-matrix of the statistical six-vertex model. The eigenvalues of the transfer matrix satisfy the famous scalar Baxter $T$-$Q$ equation, see \cite[Chapters 9, 10]{Bax82} and \cite{Bax04}. As was proved in paper \cite{Str01}, in the case $\Delta = - 1/2$ and an odd $N = 2M + 1$ this equation has a solution corresponding to an eigenvalue of the Hamiltonian $\HXXZ(\Delta)$ equal to $-3N/4$. A remarkable generalization to the case of the eight-vertex model when it describes the off-critical deformation of the $\Delta = -1/2$ six-vertex model is under active study now, see \cite{Str01a,BazMan05,BazMan06} and references therein.

It is very likely that the solution of the scalar Baxter $T$-$Q$ equation found in  the paper \cite{Str01} corresponds to the ground state of the system. Using {\sc Mathematica}, we found explicitly the eigenvectors with the eigenvalue $-3N/4$ for $N \le 17$.  Many remarkable relations between the components of the obtained eigenvectors were noticed and formulated in our paper~\cite{RazStr01a} as conjectures. It turned out, for example, that if we normalize the eigenvector so that the minimal component is $1$, then the largest component is equal to the number $A(M)$ of the alternating-sign matrices of order $M$. We refer the reader to the book \cite{Bre99} and paper \cite{BrePro99} for information on the alternating-sign matrices. Some of our conjectures are already proved \cite{KitMaiSlaTer02a,KitMaiSlaTer02b,DiFZinZub06}, and some of them are generalized to the case of different boundary conditions \cite{BatdeGNie01,RazStr01b}.

In the present paper we continue our study of the components of the ground state candidate for the XXZ spin chain with $\Delta = -1/2$ and an odd $N$. In Section 2 we observe a possible connection of some sums of components of the ground state candidate with the numbers describing the refined enumeration of the alternating-sign matrices. In Section 3 we relate those sums with the elementary symmetric polynomials in the Bethe roots, and in Section 4 we prove that these symmetric functions are related to the refined enumeration of the alternating-sign matrices.

\section{Empirical results}

In this section we consider the explicit form of the components of the ground state candidate of the Hamiltonian $\HXXZ(\Delta)$ for $\Delta = -1/2$ and an odd $N = 2M + 1$. 

Note that the shift operator $S$, defined by the equality
\[
S \, | \mu_1, \mu_2, \ldots, \mu_{N-1}, \mu_N \rangle = | \mu_2, \ldots, \mu_{N-1}, \mu_N, \mu_1 \rangle,
\]
and the reflection operator $R$, defined by the equality
\[
R \, | \mu_1, \mu_2, \ldots, \mu_{N-1}, \mu_N \rangle = | \mu_N, \mu_{N-1}, \ldots, \mu_2, \mu_1 \rangle,
\]
commute with the Hamiltonian $\HXXZ(\Delta)$, and leave the subspaces spanned by basis vectors with a fixed number of minuses invariant. As we already noted, for the case $\Delta = -1/2$ an odd $N = 2M + 1$ there are two ground states, one in the subspace with $K = M$, and another one in the subspace with $K = M + 1$. They are non-degenerate in the corresponding subspaces. Hence, each ground state is an eigenvector of the operators $S$ and $R$. Actually they are invariant with respect to the action of $S$ and $R$ that results in equality of the corresponding components. The ground state candidates under consideration are invariant with respect to the action of the operators $S$ and $R$ by construction.

Consider the ground state candidate $| \Psi \rangle$ belonging to the subspace with $K = M$. It is convenient to redenote the corresponding basis vectors as $| n_1, n_2, \ldots, n_M  \rangle$, where the integers $n_1, n_2, \ldots, n_M$ satisfy the condition $1 \le n_1 < n_2 < \ldots < n_M \le N$ and describe the positions of the minuses in the initial notation for the basis vector under consideration. The components $| \Psi \rangle$ are also marked by the integers $n_1, n_2, \ldots, n_M$. 

Consider the case $N = 11$. In this case the independent components of the ground state candidate are~\cite{RazStr01a}
\begin{align*}
& \Psi^{1,2,3,4,5} = 1, &
& \Psi^{1,3,4,5,6} = 5, &
& \Psi^{1,2,4,5,6} = 10, &
& \Psi^{1,4,5,6,7} = 11, \\[.5em]
& \Psi^{1,3,5,6,7} = 34, &
& \Psi^{1,3,4,6,7} = 41, &
& \Psi^{1,3,4,5,7} = 23, &
& \Psi^{1,2,5,6,7} = 30, \\[.5em]
& \Psi^{1,2,4,6,7} = 60, &
& \Psi^{1,5,6,7,8} = 14, &
& \Psi^{1,4,6,7,8} = 52, &
& \Psi^{1,4,5,7,8} = 73, \\[.5em]
& \Psi^{1,4,5,6,8} = 46, &
& \Psi^{1,3,6,7,8} = 75, &
& \Psi^{1,3,5,7,8} = 169, &
& \Psi^{1,3,5,6,8} = 128, \\[.5em]
& \Psi^{1,3,4,7,8} = 101, &
& \Psi^{1,2,6,7,8} = 42, &
& \Psi^{1,2,5,7,8} = 114, &
& \Psi^{1,4,7,8,9} = 81, \\[.5em]
& \Psi^{1,4,6,8,9} = 203, &
& \Psi^{1,4,6,7,9} = 174, &
& \Psi^{1,4,5,8,9} = 141, &
& \Psi^{1,3,6,8,9} = 226, \\[.5em]
& & & \Psi^{1,3,5,8,9} = 260, &
& \Psi^{1,3,5,7,9} = 429. &
\end{align*}
Here we normalize the vector in such a way that the minimal component $\Psi^{1,2,3,4,5}$ is equal to $1$. The components not given above can be obtained using the shift and reflection invariance.

In accordance with the conjecture formulated in the paper~\cite{RazStr01a} the maximal component $\Psi^{1,3,5,7,9}$ is equal to the number $A(5)$ of the alternating-sign matrices of order $5$. But this is only the tip of an iceberg.

Let us consider $\binom 5 1 = 5$ components which can be obtained from the maximal component $\Psi^{1,3,5,7,9}$ by incrementing one of its indices by $1$. We have
\begin{gather*}
\Psi^{2,3,5,7,9} = 169, \qquad \Psi^{1,4,5,7,9} = 203, \qquad \Psi^{1,3,6,7,9} = 226, \\[.5em]
\Psi^{1,3,5,8,9} = 260, \qquad \Psi^{1,3,5,7,10} = 429.
\end{gather*}
The sum of the five obtained numbers is equal to $1287$.

It is the time to look at Figure \ref{f:1} borrowed from the paper by Bressoud and Propp~\cite{BrePro99}. Let us explain its meaning.
\def \n #1{\hbox to 2.6em{\hfil #1\hfil }}
\begin{figure}[ht]
\centering \tabcolsep 0pt
\begin{tabular}{ccccccccccccc}
&&&&&&\n{1}&&&&&&\\
&&&&&\n{1}&&\n{1}&&&&&\\
&&&&\n{2}&&\n{3}&&\n{2}&&&&\\
&&&\n{7}&&\n{14}&&\n{14}&&\n{7}&&&\\
&&\n{42}&&\n{105}&&\n{135}&&\n{105}&&\n{42}&&\\
&\n{429}&&\n{1287}&&\n{2002}&&\n{2002}&&\n{1287}&&\n{429}&\\
\n{7436}&&\n{26026}&&\n{47320}&&\n{56784}&&\n{47320}&&\n{26026}&&\n{7436}
\end{tabular}
\caption{The numbers $A(M,r)$} \label{f:1}
\end{figure}

Recall that an alternating-sign matrix is a matrix with entries $1$, $0$, and $-1$ such that the $1$ and $-1$ entries alternate in each column and each row and such that the first and last nonzero entries in each row and column are $1$.
One can easily get convinced that the first column of an alternating-sign matrix contains only one entry $1$ and all other entries are zero. Denote by $A(M, r)$ the number of alternating-sign matrices of order $M$ for which the unique entry $1$ is at the $r$th position in the first column. These are the numbers that are given in the rows of Figure \ref{f:1}. One says that the numbers $A(M,r)$ give the refined enumeration of the alternating-sign matrices.

Pay our attention to the sixth row of Figure \ref{f:1}. We meet there the numbers $429$ and $1287$ already familiar to us. Let us go further and consider the components of the ground state which can be obtained from the component $\Psi^{1,3,5,7,9}$ by incrementing two its distinct indices by $1$.
The possible $\binom 5 2 = 10$ variants give
\begin{gather*}
\Psi^{2,4,5,7,9} = 128, \qquad \Psi^{2,3,6,7,9} = 101, \qquad \Psi^{2,3,5,8,9} = 114, \qquad \Psi^{2,3,5,7,10} = 203, \\
\Psi^{1,4,6,7,9} = 174, \qquad \Psi^{1,4,5,8,9} = 141, \qquad \Psi^{1,4,5,7,10} = 226, \qquad \Psi^{1,3,6,8,9} = 226, \\
\Psi^{1,3,6,7,10} = 260, \qquad \Psi^{1,3,5,7,10} = 429.
\end{gather*}
The sum of the obtained components is equal to 2002. We again obtain a number from the fifth row of Figure \ref{f:1}.

Taking into account the shift and reflection invariances, one can see that
$\binom 5 3 = 10$ components which can be obtained by incrementing three
of the five indices are numerically coincide with the corresponding components obtained by incrementing two indices, and $\binom 5 4 = 5$ components which can be obtained by incrementing four of the five indices are numerically coincide with the components obtained by incrementing one index. Furthermore, decrementing indices we obtain the same set of numbers as incrementing them.

\section{Symmetric polynomials in Bethe roots}

Let $S_K$ be the symmetric group of degree $K$. According to the Bethe ansatz \cite{Bet31} the eigenvectors of the Hamiltonian $\HXXZ(\Delta)$ for general $N$ and $K$ have the form
\begin{multline}
| \Psi \rangle = \sum_{1 \le n_1 < n_2 < \ldots < n_K \le N}
\Psi^{n_1, n_2, \ldots, n_K} | n_1, n_2, \ldots, n_K \rangle \\
= \sum_{1 \le n_1 < n_2 < \ldots < n_K \le N} \sum_{s \in S_K} A_s \, z_{s(
1)}^{n_1} z_{s(2)}^{n_2} \ldots z_{s(K)}^{n_K} | n_1, n_2, \ldots, n_K \rangle. \label{e:4}
\end{multline}
Here $z_1$, $z_2$, $\ldots$, $z_K$ are complex numbers, satisfying the
Bethe ansatz equations
\begin{equation}
z_k^N = (-1)^{K-1} \prod_{l=1}^K \left[ f(z_l, z_k) / f(z_k, z_l) \right],
\qquad k = 1, \ldots, K, \label{e:8}
\end{equation}
with
\[
f(z_1, z_2) = 1 - 2 \Delta z_2 + z_1 z_2.
\]
We call the numbers $z_1, z_2, \ldots, z_K$, satisfying (\ref{e:8}), the Bethe roots. The coefficients $A_s$ have the form
\[
A_s = \mathrm{sgn}(s)\prod_{1 \le k_1 < k_2 \le K} f(z_{s(k_2)}, z_{s(k_1)}),
\]
where $\mathrm{sgn}(s)$ is the sign of the permutation $s$. The eigenvalue of $\HXXZ(\Delta)$, corresponding to the eigenvector (\ref{e:4}), is
\begin{equation}
E = - \frac{1}{2} \Delta N + \sum_{k=1}^K (2\Delta - z_k - z_k^{-1}). \label{e:5}
\end{equation}

It is convenient for our purposes to introduce the notation
\[
B_s^{n_1,n_2,\ldots,n_K} = A_s z_{s(1)}^{n_1} z_{s(2)}^{n_2} \ldots z_{s(K)}^{n_K}.
\]
Using this notation, we can write
\[
\Psi^{n_1, n_2, \ldots, n_K} = \sum_{s \in S_K} B_s^{n_1, n_2, \ldots, n_K}.
\]

Return again to the case of an odd $N$. Consider an eigenvector $| \Psi \rangle$ belonging to the subspace with $K = M$, not necessary a ground state. Denote by $F$ the component $\Psi^{1,3,\ldots,2M-1}$, so that
\[
F = \sum_{s \in S_M} B_s^{1,3,\ldots,2M-1}.
\]
Incrementing an index of $\Psi^{1,3,\ldots,2M-1}$ by $1$ we obtain $M$ components
\[
F_m = \Psi^{1, 3, \ldots, 2m-3, 2m, 2m+1, \ldots, 2M-1}.
\]
It is clear that
\[
F_m  = \sum_{s \in S_M} z_{s(m)} B_s^{1, 3, \ldots, 2m-3, 2m-1, 2m+1, \ldots,2M-1},
\]
and after summation over $m$ we have 
\begin{multline*}
\sum_{m=1}^M F_m = \sum_{m=1}^M \sum_{s \in S_M} z_{s(m)} B_s^{1, 3,
\ldots, 2m-3, 2m-1, 2m+1, \ldots, 2M-1} \\
= \sum_{s \in S_M} \left( \sum_{m=1}^M  z_{s(m)} \right)
B_s^{1, 3,\ldots, 2m-3, 2m-1, 2m+1, \ldots, 2M-1} \\ = \left( \sum_{m=1}^M  z_m
\right) \sum_{s \in S_M} B_s^{1, 3, \ldots, 2m-3, 2m-1, 2m+1, \ldots, 2M-1},
\end{multline*}
where we use the evident equality
\[
\sum_{m=1}^M z_{s(m)} = \sum_{m=1}^M z_m.
\]
Thus, for the first symmetric polynomial \cite{Lan02} in the Bethe roots we have the expression
\[
e_1(z_1, \ldots, z_M) = \sum_{m=1}^M z_m = \frac{1}{F} \sum_{k=1}^M F_m.
\]

Denote now by $F_{m_1, \ldots, m_r}$ the component which is obtained from $\Psi^{1, 3, \ldots, 2M-1}$ by incrementing the indices at the positions $m_1, \ldots, m_r$ by 1. Here we assume that $1 \le m_1 < \ldots < m_r \le M$. In a similar way as above, using the equality
\[
\sum_{1 \le m_1 < \ldots < m_r \le M} z_{s(m_1)} \ldots z_{s(m_1)} = \sum_{1 \le m_1 < \ldots < m_r \le M} z_{m_1} \ldots z_{m_r},
\]
we see that the $r$th elementary symmetric polynomial in the Bethe roots is related to the components of the eigenvector $| \Psi \rangle$ as
\[
e_r(z_1, \ldots, z_M) = \sum_{1 \le m_1 < m_2 < \ldots < m_r \le M} z_{m_1} z_{m_2} \ldots z_{m_r} = \frac{1}{F} \sum_{1 \le m_1 < m_2 < \ldots < m_r \le M} F_{m_1,
m_2, \ldots, m_r}.
\]

In the next section for the case of $\Delta = -1/2$ and an odd $N$ we show that for the considered state with the eigenvalue $-3N/4$ the elementary symmetric polynomials in the Bethe roots coincide with the numbers giving the refined enumeration of the alternating-sign matrices.

\section{Relation to refined enumeration of alternating-sign matrices}

It is well known that the XXZ spin chain is closely related to the six-vertex model, see, for example, the monograph \cite{Bax82}. To describe this relation, let us parametrize the Boltzmann weights of the six-vertex model in the following way
\[
a(u) = \sigma(\tau u), \qquad b(u) = \sigma(\tau^{-1} u), \qquad c(u) = \sigma(\tau^2),
\]
where $\tau$ is a fixed parameter, $u$ is a free parameter, and we use the convenient notation
\[
\sigma(x) = x - x^{-1}
\]
introduced by Kuperberg. One can show that the Hamiltonian $\HXXZ(\Delta)$ is connected with the commuting family $T(u)$ of transfer matrices of the six-vertex model by the equality
\begin{equation}
\left[T^{-1}(u) \frac{\rmd T(u)}{\rmd u} \right]_{u=\tau} = - \frac{2}{\tau(\tau^2 - \tau^{-2})} \left[ \HXXZ(\Delta(\tau)) - \frac{N}{4} (\tau^2 + \tau^{-2}) \right], \label{e:17}
\end{equation}
where
\begin{equation}
\Delta(\tau) = (\tau^2 + \tau^{-2})/2. \label{e:3}
\end{equation}
Therefore, one can diagonalize the Hamiltonian $\HXXZ(\Delta)$ with $\Delta$ given by (\ref{e:3}) and the family $T(u)$ simultaneously. 

It appears that the vector $| \Psi \rangle$, defined by relation (\ref{e:4}), is an eigenvector of $T(u)$ with  the eigenvalue
\begin{equation}
\lambda(u) = a^N(u) U(u, z_1) \ldots U(u, z_K) + b^N(u) V(u, z_1) \ldots V(u, z_K), \label{e:2}
\end{equation}
where
\begin{gather*}
U(u, z) = \frac{a(u)b(u) + (c^2(u) - b^2(u))z}{a^2(u) - a(u) b(u) z}, \qquad
V(u, z) = \frac{a^2(u) - c^2(u) - a(u) b(u) z}{a(u) b(u) - b(u)^2 z},
\end{gather*}
see, for example, the paper \cite{Lie67} and the monograph \cite{Bax82}. It is convenient to introduce the complex numbers $u_k$, $k = 1, \ldots, K$, related to the Bethe roots $z_k$ by the equality
\begin{equation}
z_k = \frac{a(u_k)}{b(u_k)} = \frac{\sigma(\tau u_k)}{\sigma(\tau^{-1} u_k)} = \frac{\tau^2 u_k^2 - 1}{u_k^2 - \tau^2}. \label{e:9}
\end{equation}
It is clear that the numbers $u_k$ are defined up to a sign. One easily obtains
\[
U(u, z_k) = \frac{\sigma(\tau^{-2} (u u_k^{-1}))}{\sigma(u u_k^{-1})}, \qquad V(u, z_k) = \frac{\sigma(\tau^2 (u u_k^{-1}))}{\sigma(u u_k^{-1})}.
\]
Denoting\footnote{Actually the function $\xi(u)$ is defined up to a factor. We define it in such a way that the coefficient at the maximal degree of $u$ is equal to 1.}
\begin{equation}
\xi(u) = \prod_{k=1}^K [u_k \, \sigma(u u_k^{-1})], \label{e:13}
\end{equation}
we rewrite equality (\ref{e:2}) as
\begin{equation}
\lambda(u) \xi(u) = \sigma^N(\tau u) \xi(\tau^{-2} u) + \sigma^N(\tau^{-1} u) \xi(\tau^2 u). \label{e:7}
\end{equation}
Treating $\xi(u)$ as eigenvalues of a commuting family of the operators $Q(u)$, we come to the equality
\[
T(u) Q(u) = \sigma^N(\tau u) Q(\tau^{-2} u) + \sigma^N(\tau^{-1} u) Q(\tau^2 u),
\]
which is the famous Baxter $T$-$Q$ equation. The equation (\ref{e:7}) is called the scalar Baxter $T$-$Q$ equation. One can solve equation (\ref{e:7}) instead of solving the Bethe ansatz equations (\ref{e:8}). The corresponding Bethe roots $z_k$ can be recovered from the roots $u_k$ of the Laurent polynomial $\xi(u)$ using relation (\ref{e:9}).

Assume now that $\tau = \exp(\rmi \pi/3)$ that gives $\Delta = -1/2$.
It appears that for an odd $N = 2M + 1$ the Bethe equations in the subspace with $K = M$ has a solution corresponding to
\[
\lambda(u) = [a(u) + b(u)]^N = \left[ (\tau + \tau^{-1})(u - u^{-1}) \right]^N = \sigma^N(u).
\]
Here we use the identity
\[
\tau + \tau^{-1} = 1.
\]
To find the solution in question, one observes that defining
\[
\varphi(u) = \sigma^N(u) \xi(u)
\]
and using the equality
\[
\sigma(\tau^3 x) = - \sigma(x),
\]
one can rewrite equation (\ref{e:7}) as
\begin{equation}
\varphi(u) + \varphi(\tau^2 u) + \varphi(\tau^4 u) = 0. \label{e:10}
\end{equation}

Let us suppose that the solution under consideration corresponds to a ground state of the system in the subspace with $K = M$. As was proved in the paper \cite{YanYan66}, in this case one can number the Bethe roots in such a way that
\begin{equation}
z_m^{-1} = z_{M-m+1}. \label{e:6}
\end{equation}
One can chose the corresponding numbers $u_m$ so that
\[
u_m^{-1} = - u_{M-m+1}.
\]
In this case the function $\xi(u)$ satisfies the equality
\[
\xi(u^{-1}) = \xi(u)
\]
that gives
\begin{equation}
\varphi(u^{-1}) = - \varphi(u). \label{e:11}
\end{equation}
It is also clear that $\varphi(u)$ should be a centered Laurent polynomial in $u$ of degree $6M + 2$ and that
\begin{equation}
\varphi(-u) = (-1)^{M+1} \varphi(u). \label{e:12}
\end{equation}
Finally note that the Laurent polynomial $\varphi(u)$ should be divisible by $\sigma^{2M+1}(u)$ by construction.

It was shown in the paper \cite{Str01} that up to a constant factor there is a unique solution of equation (\ref{e:10}) which is a centered Laurent polynomial in $u$ of degree $6M + 2$, satisfies relations (\ref{e:11}) and (\ref{e:12}), and is divisible by $\sigma^{2M+1}(u)$. This polynomial has the form
\begin{equation}
\varphi(u) = \binom{M - 1/3}{M}^{-1} \sum_{m=0}^M \binom{M - 1/3}{m} \binom{M + 1/3}{M - m} \sigma(u^{1 - 3M + 6m}). \label{e:15}
\end{equation}
We chose the normalization in accordance with the definition (\ref{e:13}) of $\xi(u)$.

Let us now relate the Laurent polynomial $\varphi(u)$ to elementary symmetric polynomials in the Bethe roots.  Define the variable
\[
z = \frac{a(u)}{b(u)} = \frac{\sigma(\tau u)}{\sigma(\tau^{-1} u)} = \frac{\tau^2 u^2 - 1}{u^2 - \tau^2},
\]
then, using relation (\ref{e:9}), one obtains
\begin{equation}
z - z_m = \frac{\sigma(\tau^2) \sigma(u u_m^{-1})}{\sigma(\tau^{-1} u) \sigma(\tau u_m^{-1})}. \label{e:14}
\end{equation}
Consider the function
\[
\chi(z) = \prod_{m = 1}^M (z - z_m) = \sum_{r=0}^M (-1)^r e_r(z_1, \ldots, z_M) z^{M-r},
\]
where we assume that $e_0(z_1, \ldots, z_M) = 1$. Using (\ref{e:14}), we come to the equality
\begin{equation}
\chi(z) = \prod_{m = 1}^M (z - z_m) = \frac{\sigma^M(\tau^2) \sigma^{2M + 1}(\tau) \varphi(u)}{\sigma^M(\tau^{-1} u) \sigma^{2M + 1}(u) \varphi(\tau)}. \label{e:16}
\end{equation}

It follows from the explicit representation (\ref{e:15}) for the function $\varphi(u)$ that it satisfies the second order differential equation
\[
u \frac{\rmd}{\rmd u} \left( u \frac{\rmd \varphi(u)}{\rmd u} \right) - 6 M \frac{u^6 + 1}{u^6 - 1} u \frac{\rmd \varphi(u)}{\rmd u} + (3M + 1)(3M
- 1) \varphi(u) = 0.
\] 
Now one can show that the function $\chi(z)$, related to $\varphi(u)$ by equality (\ref{e:16}), satisfies the differential equation
\[
z(z + 1) \frac{\rmd^2 \chi(z)}{\rmd z^2} + 2(z - M) \frac{\rmd \chi(z)}{\rmd z} - M(M+1) \chi(z) = 0.
\]
Using this equation, we come to the recursive relation for the elementary symmetric polynomials in the Bethe roots,
\[
(2M - r + 1) r \, e_r(z_1, \ldots, z_M) = (M - r + 1)(M + r) \, e_{r-1}(z_1, \ldots, z_M).
\]
Comparing it with the recursive relation for the numbers giving the refined enumeration of the alternating-sign matrices
\[
(2M - r - 1) r \, A(M, r + 1) = (M - r)(M + r - 1) \, A(M, r),
\]
see, for example, \cite{Str05}, and using the initial condition $e_0(z_1, \ldots, z_M) = 1$, we see that
\[
e_r(z_1, \ldots, z_M) = \frac{1}{A(M)} A(M + 1, r + 1).
\]
Hence, if our conjecture, stating that if we normalize the ground state candidate so that the minimal component is equal to $1$, then the largest component is $A(M)$ \cite{RazStr01a}, we come to the relation
\[
\sum_{1 \le m_1 < m_2 < \ldots < m_r \le M} F_{m_1, m_2, \ldots, m_r} = A(M + 1, r+1).
\]

For the simplest symmetric polynomial in the Bethe roots, $e_1(z_1, \ldots, z_M)$, we have
\[
e_1(z_1, \ldots, z_M) = \sum_{m=1}^M z_m = (M+1)/2.
\]
Taking into account that in the case under consideration
\[
\sum_{m=1}^M z_m^{-1} = \sum_{m=1}^M z_m,
\]
we obtain from equality (\ref{e:5}) that the eigenvalue of $\HXXZ(\Delta)$, corresponding to the discussed ground state candidate, is equal to $- 3 N / 4$. Certainly, this result can be obtained in a more direct way, starting from relation (\ref{e:17}).

\section{Discussion}

In the present paper we considered the properties of the most probable ground state candidate for the XXZ spin chain with the anisotropy parameter $\Delta$ equal to $-1/2$ and an odd number of sites $N$. We related some combinations of the components of the considered state to the elementary symmetric polynomials in the corresponding Bethe roots. Then we proved that those polynomials coincide with the numbers giving the refined enumeration of the alternating-sign matrices.

The components $\Psi^{n_1, \ldots, n_K}$, defined via equality (\ref{e:4}) are antisymmetric polynomials in the Bethe roots $z_1, \ldots, z_K$. Representing them as
\[
\Psi^{n_1, \ldots, n_K} = \left( \prod_{1 \le k_1 < k_2 \le K} (z_{k_1} - z_{k_2}) \right) \Phi^{n_1, \ldots, n_K},
\]
we introduce the quantities $\Phi^{n_1, \ldots, n_K}$ which are symmetric polynomials in the Bethe roots. Therefore, the components of an eigenvector of the Hamiltonian $\HXXZ$, given by the Bethe ansatz (\ref{e:4}), can be expressed, in principle, via elementary symmetric polynomials in the Bethe roots. Thus, the components of our ground state candidate can be expressed via the numbers giving the refined enumeration of the alternating-sign matrices.

The authors of the paper \cite{deGBatNieMit01}, starting from relation (\ref{e:15}), found, in particular, a representation of the elementary symmetric polynomials in the Bethe roots via some sums of products of binomial coefficients. 

Special solutions to the Baxter $T$-$Q$ equation can be found for different boundary conditions \cite{FriStrZag00,FriStrZag01}, and our results can be generalized to those cases.

\subsection*{Acknowledgment}

The work was supported in part by the Russian Foundation for Basic Research under grant \# 04--01--352. It is a pleasure to thank P. Di Francesco, V. Pasquier, P. Zinn--Justin and J.--B. Zuber for interesting discussions. We wish to acknowledge the warm hospitality of the Laboratoire de Physique Theoriqu\'e et Mod\`eles Statistiques of Universit\'e Paris-Sud, where this work was finished.

\end{document}